# Performance Comparison of two novel Relay-Assisted Hybrid FSO / RF Communication Systems

*Mohammad Ali Amirabadi[1],*

[1] *School of Electrical Engineering, Iran University of Science and Technology, Tehran, Iran*
✉ *E-mail: m_amirabadi@elec.iust.ac.ir*

**Abstract:** In this manuscript, two novel multi-hop relay-assisted hybrid Free Space Optical / Radio Frequency (FSO / RF) communication systems are presented and compared. In these structures, RF and FSO links, at each hop, are parallel and send data simultaneously. This is the first time that in a multihop hybrid FSO / RF structure, Detect and Forward protocol is used. In the first structure, at each hop, received signals with higher Signal to Noise Ratio (SNR) is selected. But in the second structure, at each hop, received FSO and RF signals are separately detected and forwarded and selection is done only at the last hop. Considering FSO link in Negative Exponential atmospheric turbulence and RF link in Rayleigh fading, for the first time, closed-form expressions are derived for Outage Probability ($P_{out}$) and Bit Error Rate ($BER$) of the proposed structures. MATLAB simulations are provided to verify derived expressions. The main motivation of this work is to answer this question that how much is the difference of selection at each hop and selection at the last hop. Results indicate that the structure with selection at each hop has better performance than the structure with selection at the last hop. At different target Outage Probability, selection at the last hop consumes about two times ($\sim$**3dB**) more power than selection at each hop. Both structures are particularly suitable for long-range communications. However, selection at each hop, in the cost of more complexity, is recommended for applications which have problem with supplying the required power for communication.

## 1 Introduction

FSO communication systems are considered as an appropriate alternative for traditional RF communication systems. FSO system has large bandwidth compared with RF systems. In addition, because of a very narrow beam, FSO is highly secure and contains no interference. Besides these advantages, constraints such as high sensitivity to atmospheric turbulence and weather conditions, severely limits FSO practical applications [1].

Various statistical distributions have been used to model atmospheric turbulence effects. Lognormal [2], Gamma-Gamma [3] and Negative Exponential [4] distributions are respectively used to model weak, moderate to strong and saturate atmospheric turbulence regimes.

Effects of atmospheric turbulences on FSO and RF links are not the same, in the sense that when atmospheric turbulences cause outage in one of the FSO and RF links, the other link remains available. For example, severe rain does not affect performance of FSO link, but degrades the performance of RF link. Therefore combination of FSO and RF links can significantly improve the performance and reliability of the system [5].

Hybrid FSO / RF systems are available in series [6,7] and parallel [8,9] structures. In parallel structure, FSO and RF links are parallel and send data simultaneously [10] or by use of a switch [11]. In simultaneous data transmission, both FSO and RF links are always active, but in switching method, FSO link is always active and RF link acts as a backup, i.e. when the received SNR comes down below a threshold level, RF link starts transmitting data. Series structure has lower power consumption, but suffers performance degradation due to frequent switches when atmospheric turbulence gets worse [12].

The so-called relay assisted hybrid FSO / RF systems improve both performance and capacity of the system. Different relay protocols have been studied for relay assisted systems; among them Amplify and Forward [13], due to its simplicity is mostly used. In this protocol, according to the existence or missing of Channel State Information (CSI), the received signal is amplified and forwarded by adaptive or fixed gain [14,15]. Also the case of outdated CSI is studied in [16]. Amplify and Forward has high power consumption and amplifies noise. When CSI is missed, amplification gain should be adjusted according to the worst case scenario, therefore, it has power loss. It is better to use other protocols such as Decode and Forward [17], and Detect and Forward [18].

Works done about relay-assisted hybrid FSO / RF systems, mostly considered one hop [19, 20] and two hop [21-24] structures. Few works have been done about multihop hybrid FSO / RF systems [25, 26]. To the best of the authors' knowledge, this the first time a multi hop relay-assisted hybrid FSO / RF system with parallel data transmission on each hop, is being investigated at saturate atmospheric turbulence.

In this paper two novel models are presented for multihop hybrid FSO / RF system. At each hop of the proposed structures, FSO and RF links are parallel and send data simultaneously. Detect and forward protocol is used. At each relay input the first structure, received signal with higher SNR is selected. But in the second structure, the received signal at each relay input are detected and forwarded separately, and selection is done only at the destination. To the best of the authors' knowledge, this is the first time that Detect and Forward protocol is used for multihop hybrid FSO / RF system. Also it is the first time that BER of a multihop hybrid FSO/RF is investigated. Considering FSO link in Negative Exponential atmospheric turbulence and RF link in Rayleigh fading, for the first time, closed-form expressions are derived for $BER$ and $P_{out}$ of the proposed structures. Derived expressions are validated trough MATLAB simulations. In proposed structures, combination of FSO and RF links significantly increases link reliability and accessibility. In addition, multihop relaying reduces total power consumption and increases capacity of the system. Also detect and forward relaying used in this work has low cost and easy implementation. The proposed structures are particularly suitable for long-range communications.

The rest of this work is organized as follows. In Section 2, system model, $P_{out}$ and $BER$ of the structure with selection at each hop are discussed. System model, $P_{out}$ and $BER$ of the structure with selection at the destination are expressed in section 3. In Section 4 analytical and simulation results are compared and discussed. Section 5 is the conclusion of this work.

## 2 The structure with selection at each hop

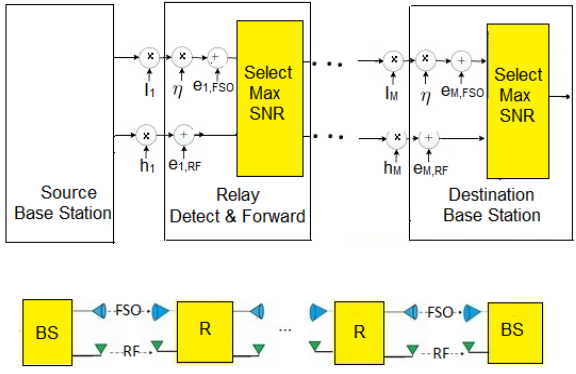

**Fig. 1:** The structure with selection at each hop

### 2.1 System Model

The proposed multihop hybrid FSO / RF system is presented in Fig. 1. Assuming $x$ as the generated signal at source Base Station; two copies of this signal are transmitted through FSO and RF links. Between received FSO and RF signals at the first relay, signal with higher SNR is detected, then two copies of this signal are forwarded through RF and FSO links. The same trend continues at each hop until reaching the destination Base Station.
Received FSO and RF signals at the $i$-th; $i = 1.2.\ldots.M$ relay are as:

$$y_{FSO.i} = \eta I_i d_{i-1} + e_{FSO.i}, \quad (1)$$
$$y_{RF.i} = h_i d_{i-1} + e_{RF.i}, \quad (2)$$

where $d_{i-1}$ is the detected signal at the $(i-1)-th$ relay, $h_i$ is fading coefficient of $i$-th hop, $I_i$ is atmospheric turbulence intensity of $i$-th hop, $e_{FSO.i}$ is Additive White Gaussian Noise (AWGN) with zero mean and $\sigma_{FSO}^2$ variance at the input of $i-th$ FSO receiver, $e_{RF.i}$, is AWGN with zero mean and $\sigma_{RF}^2$ variance at the input of $i-th$ RF receiver, and $\eta$ is photo detector responsivity. At each hop, between received FSO and RF signals, signal with the highest SNR is selected for detection, i.e. $\gamma_i = max(\gamma_{FSO.i}.\gamma_{RF.i})$. Assuming independence of FSO and RF links, the Cumulative Distribution Function (CDF) of $\gamma_i$ becomes as follows:

$$F_{\gamma_i}(\gamma) = Pr(max(\gamma_{FSO.i}.\gamma_{RF.i}) \le \gamma) = Pr(\gamma_{FSO.i} \le \gamma.\gamma_{RF.i} \le \gamma) = F_{\gamma_{FSO.i}}(\gamma)F_{\gamma_{RF.i}}(\gamma), \quad (3)$$

where $\gamma_{FSO.i}$ is SNR at the input of $i-th$ FSO receiver and $\gamma_{RF.i}$ is SNR at the input of $i-th$ RF receiver.

### 2.2 Outage Probability

Outage occurs when the SNR comes below a threshold level, i.e. $\gamma \le \gamma_{th}$. Assuming that the detection done at each relay is independent from the detection done at other relays, $P_{out}$ of the first proposed structure will be defined as follows [26]:

$$P_{out}(\gamma_{th}) = 1 - P_{ava}(\gamma_{th}) = 1 - \prod_{i=1}^{M} P_{ava.i}(\gamma_{th}) = 1 - \prod_{i=1}^{M}(1 - P_{out.i}(\gamma_{th})), \quad (4)$$

where $P_{ava} = 1 - P_{out}$ is probability of link availability. So, given that $P_{out}(\gamma_{th}) = F_\gamma(\gamma_{th})$, and that FSO link atmospheric turbulence and RF link fading are independent and identically distributed, the above statement becomes equal to:

$$P_{out}(\gamma_{th}) = 1 - \left(1 - F_{\gamma_i}(\gamma_{th})\right)^M = 1 - \left(1 - F_{\gamma_{RF.i}}(\gamma_{th})F_{\gamma_{FSO.i}}(\gamma_{th})\right)^M. \quad (5)$$

Therefore, in order to obtain $P_{out}$ of the first proposed structure, it is sufficient to calculate CDF of FSO and RF links, separately and then substitute them at (5).
FSO link is described by Negative Exponential atmospheric turbulence with $1/\lambda$ mean and $1/\lambda^2$ variance. The probability density function (pdf) of $I_i$ is as follows:

$$f_{I_i}(I_i) = \lambda e^{-\lambda I_i}. \quad (6)$$

According to (1), instantaneous SNR at the $i-th$ FSO receiver is as $\gamma_{FSO.i} = \frac{\eta^2 I_i^2}{\sigma_{FSO}^2} = \bar{\gamma}_{FSO} I_i^2$, where $\bar{\gamma}_{FSO}$ is average SNR. Using [27] and (6), the pdf and CDF of $\gamma_{FSO.i}$ are equal to:

$$f_{\gamma_{FSO.i}}(\gamma) = \frac{\lambda}{2\sqrt{\bar{\gamma}_{FSO}\gamma}} e^{-\lambda\sqrt{\frac{\gamma}{\bar{\gamma}_{FSO}}}}, \quad (7)$$

$$F_{\gamma_{FSO.i}}(\gamma) = 1 - e^{-\lambda\sqrt{\frac{\gamma}{\bar{\gamma}_{FSO}}}}.$$

According to (2), instantaneous SNR at the $i-th$ RF receiver is as $\gamma_{RF.i} = \frac{h_i^2}{\sigma_{RF}^2} = \bar{\gamma}_{RF} h_i^2$, where $\bar{\gamma}_{RF}$ is average SNR. The pdf and CDF of $\gamma_{RF.i}$ are equal to:

$$f_{\gamma_{RF.i}}(\gamma) = \frac{1}{\bar{\gamma}_{RF}} e^{-\frac{\gamma}{\bar{\gamma}_{RF}}}, \quad (8)$$

$$F_{\gamma_{RF.i}}(\gamma) = 1 - e^{-\frac{\gamma}{\bar{\gamma}_{RF}}}. \quad (9)$$

Substituting (8) and (10) into (5), $P_{out}$ of the proposed structure is as follows:

$$P_{out}(\gamma_{th}) = 1 - \left(1 - \left(1 - e^{-\lambda\sqrt{\frac{\gamma_{th}}{\bar{\gamma}_{FSO}}}}\right)\left(1 - e^{-\frac{\gamma_{th}}{\bar{\gamma}_{RF}}}\right)\right)^M. \quad (10)$$

Substituting binomial expansion of $\left[1 - \left(1 - e^{-\lambda\sqrt{\frac{\gamma_{th}}{\bar{\gamma}_{FSO}}}}\right)\left(1 - e^{-\frac{\gamma_{th}}{\bar{\gamma}_{RF}}}\right)\right]^M$ as $\sum_{k=0}^{M}\binom{M}{k}(-1)^k\left(1 - e^{-\lambda\sqrt{\frac{\gamma_{th}}{\bar{\gamma}_{FSO}}}}\right)^k\left(1 - e^{-\frac{\gamma_{th}}{\bar{\gamma}_{RF}}}\right)^k, \quad (11)$
becomes equal to:

$$P_{out}(\gamma_{th}) = 1 - \sum_{k=0}^{M}\binom{M}{k}(-1)^k\left(1 - e^{-\lambda\sqrt{\frac{\gamma_{th}}{\bar{\gamma}_{FSO}}}}\right)^k\left(1 - e^{-\frac{\gamma_{th}}{\bar{\gamma}_{RF}}}\right)^k. \quad (11)$$

Substituting binomial expansion of $\left(1 - e^{-\frac{\gamma_{th}}{\bar{\gamma}_{RF}}}\right)^k$ and $\left(1 - e^{-\lambda\sqrt{\frac{\gamma_{th}}{\bar{\gamma}_{FSO}}}}\right)^k$, respectively as $\sum_{v=0}^{k}\binom{k}{v}(-1)^v e^{-\frac{v\gamma_{th}}{\bar{\gamma}_{RF}}}$ and $\sum_{u=0}^{k}\binom{k}{u}(-1)^u e^{-\lambda u\sqrt{\frac{\gamma_{th}}{\bar{\gamma}_{FSO}}}}$, $P_{out}$ of the proposed structure becomes equal to:

$$P_{out}(\gamma_{th}) = 1 - \sum_{k=0}^{M}\sum_{v=0}^{k}\sum_{u=0}^{k}\binom{M}{k}\binom{k}{v}\binom{k}{u}(-1)^{k+v+u}e^{-\frac{v\gamma_{th}}{\bar{\gamma}_{RF}}}e^{-\lambda u\sqrt{\frac{\gamma_{th}}{\bar{\gamma}_{FSO}}}}. \quad (12)$$

$P_{out}$ of the proposed structure is related exponentially to $1/\bar{\gamma}_{RF}$ and $1/\sqrt{\bar{\gamma}_{FSO}}$, therefore in order to improve performance of the system, it is better to increase $\bar{\gamma}_{RF}$, rather than $\bar{\gamma}_{FSO}$. In the sense that increasing the transmitted power in RF link brings more improvement than that of FSO link.

### 2.3 Bit Error Rate

Though performance of MPSK modulations is better than DPSK, but DPSK does not need the carrier phase estimation circuit and therefore, its receiver has less complexity. Given that $F_\gamma(\gamma) = P_{out}(\gamma)$, BER for DPSK modulation is obtained from the following equation [28]:

$$P_e = \frac{1}{2}\int_0^\infty e^{-\gamma} F_\gamma(\gamma) d\gamma = \frac{1}{2}\int_0^\infty e^{-\gamma} P_{out}(\gamma) d\gamma. \quad (13)$$

Substituting (13) in (14), BER of DPSK modulation of the proposed structure becomes equal to:

$$P_e = \frac{1}{2}\int_0^\infty e^{-\gamma}\left(1 - \sum_{k=0}^{M}\sum_{v=0}^{k}\sum_{u=0}^{k}\binom{M}{k}\binom{k}{v}\binom{k}{u}(-1)^{k+v+u}e^{-\frac{v\gamma}{\bar{\gamma}_{RF}}}e^{-\lambda u\sqrt{\frac{\gamma}{\bar{\gamma}_{FSO}}}}\right)d\gamma. \quad (14)$$

Substituting the Meijer-G equivalent of $e^{-\lambda u\sqrt{\gamma/\bar{\gamma}_{FSO}}}$ as $\frac{1}{\sqrt{\pi}}G_{0,2}^{2,0}\left(\frac{\lambda^2 u^2 \gamma}{4\bar{\gamma}_{FSO}}\Big|_{0,\frac{1}{2}}^{-}\right)$ [29, Eq. 07.34.03.1081.01] and using [29, Eq. 07.34.21.0088.01] BER of DPSK modulation of the proposed structure becomes equal to:

$$P_e = \frac{1}{2}\Bigg(1 - \sum_{k=0}^{M}\sum_{v=0}^{k}\sum_{u=0}^{k}\binom{M}{k}\binom{k}{v}\binom{k}{u} \times \frac{(-1)^{k+v+u}}{\sqrt{\pi}\left(1+\frac{v}{\bar{\gamma}_{RF}}\right)}G_{1,2}^{2,1}\left(\frac{\lambda^2 u^2}{4\bar{\gamma}_{FSO}\left(1+\frac{v}{\bar{\gamma}_{RF}}\right)}\Big|_{0,\frac{1}{2}}^{0}\right)\Bigg). \quad (15)$$

## 3 The structure with selection at the destination.

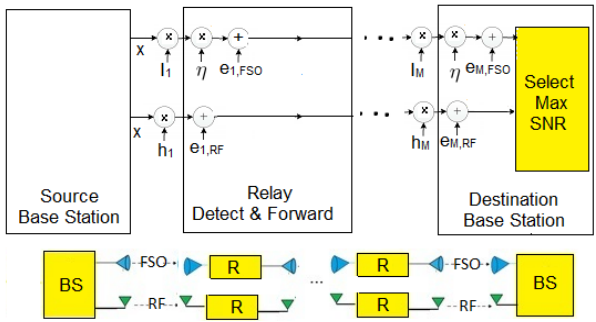

**Fig. 2**: The structure with selection at the destination.

### 3.1 System model

In Fig. 2, a relay-assisted hybrid FSO / RF communication system with M series hops is considered, in which FSO and RF links are parallel and send data simultaneously and separately. Assuming $x$ as the generated signal at the source Base Station, two copies of the signal are transmitted through FSO and RF links. The received FSO and RF signals are then detected and forwarded separately. This procedure continues until reaching the destination Base Station. At the destination, between received FSO and RF signals, one with higher SNR is used for detection.

The received FSO and RF signals at $i-th$ relay are as follows:

$$y_{FSO.i} = \eta I_i d_{FSO.i-1} + e_{FSO.i}, \quad (17)$$
$$y_{RF.i} = h_i d_{RF.i-1} + e_{RF.i}. \quad (18)$$

where $d_{RF.i-1}$ and $d_{FSO.i-1}$ are respectively the detected FSO and RF signals at $(i-1)-th$ relay.

### 3.2 Outage Probability

Assuming independent FSO and RF links and assuming independent detection at each relay, $P_{out}$ of the second proposed structure is defined as follows:

$$P_{out}(\gamma_{th}) = P_{out.FSO}(\gamma_{th})P_{out.RF}(\gamma_{th}) = \left(1 - P_{ava.FSO}(\gamma_{th})\right)\left(1 - P_{ava.RF}(\gamma_{th})\right) = \left(1 - \prod_{i=1}^{M}\left(1 - P_{out.RF.i}(\gamma_{th})\right)\right)\left(1 - \prod_{i=1}^{M}\left(1 - P_{out.FSO.i}(\gamma_{th})\right)\right). \quad (19)$$

Given that $P_{out}(\gamma_{th}) = F_\gamma(\gamma_{th})$, and independent identically distributed FSO and RF links, (19) becomes equal to:

$$P_{out}(\gamma_{th}) = \left(1 - \left(1 - F_{\gamma_{RF.i}}(\gamma_{th})\right)^M\right)\left(1 - \left(1 - F_{\gamma_{FSO.i}}(\gamma_{th})\right)^M\right). \quad (20)$$

Substituting (8) and (10) into (21), $P_{out}$ of the second proposed structure becomes equal to:

$$P_{out}(\gamma_{th}) = 1 - e^{-\lambda M\sqrt{\frac{\gamma_{th}}{\bar{\gamma}_{FSO}}}} - e^{-\frac{M\gamma_{th}}{\bar{\gamma}_{RF}}} + e^{-\frac{M\gamma_{th}}{\bar{\gamma}_{RF}}}e^{-\lambda M\sqrt{\frac{\gamma_{th}}{\bar{\gamma}_{FSO}}}}. \quad (21)$$

### 3.3 Bit Error Rate

Substituting (21) in (14), and inserting Meijer-G equivalent of $e^{-\lambda M\sqrt{\gamma/\bar{\gamma}_{FSO}}}$ in it, BER of DPSK modulation of the second proposed structure becomes as follows:

$$P_e = \frac{1}{2}\int_0^\infty e^{-\gamma}\left(1 - \frac{1}{\sqrt{\pi}}G_{0,2}^{2,0}\left(\frac{\lambda^2 M^2 \gamma}{4\bar{\gamma}_{FSO}}\Big|_{0,\frac{1}{2}}^{0}\right) - e^{-\frac{M\gamma}{\bar{\gamma}_{RF}}} + \frac{1}{\sqrt{\pi}}e^{-\frac{M\gamma}{\bar{\gamma}_{RF}}}G_{0,2}^{2,0}\left(\frac{\lambda^2 M^2 \gamma}{4\bar{\gamma}_{FSO}}\Big|_{0,\frac{1}{2}}^{0}\right)\right)d\gamma. \quad (22)$$

Using [29, Eq. 07.34.21.0088.01] BER of DPSK modulation of the second structure becomes equal to:

$$P_e = \frac{1}{2}\Bigg(1 - \frac{1}{\sqrt{\pi}}G_{1,2}^{2,1}\left(\frac{\lambda^2 M^2}{4\bar{\gamma}_{FSO}}\Big|_{0,\frac{1}{2}}^{0}\right) - \frac{1}{1+\frac{M}{\bar{\gamma}_{RF}}} + \frac{1}{\sqrt{\pi}}\frac{1}{1+\frac{M}{\bar{\gamma}_{RF}}}G_{1,2}^{2,1}\left(\frac{\lambda^2 M^2}{4\bar{\gamma}_{FSO}\left(1+\frac{M}{\bar{\gamma}_{RF}}\right)}\Big|_{0,\frac{1}{2}}^{0}\right)\Bigg). \quad (23)$$

## 4 Comparison between analytical and simulation results

In this section mathematical analysis and MATLAB simulation results are compared. FSO and RF links are described by Negative Exponential atmospheric turbulence and Rayleigh fading, respectively. Performances of the proposed structures are investigated at various variances of Negative Exponential atmospheric turbulence ($1/\lambda^2$) and various numbers of hops (M). FSO and RF links have equal average SNR ($\bar{\gamma}_{FSO} = \bar{\gamma}_{RF} = \gamma_{avg}$). $\gamma_{th}$ is outage threshold SNR.

In Fig. 3, Outage Probability of the proposed structures is plotted in terms of average SNR for various number of hops for Negative Exponential atmospheric turbulence with unit variance ($\lambda = 1$) and $\gamma_{th} = 10dB$. As can be seen, the performance of the proposed structures degrades by increasing the number of hops. Because according to the definition of $P_{out}$, and due to use of series relaying structure, outage occurrence in the proposed structures is caused by the outage even in one hop, therefore adding number of hops increases $P_{out}$. In fact, multihop structure replaces high power consumption long–range link with some low power consumption short-range links.

As can be seen, the structure with selection at each hop has better performance than the other structure better than the other structure, because in the structure with selection at each hop, between the two received FSO and RF signals, the one with higher SNR is used for detection at each hop but at the structure with selection at the destination, each of the received FSO and RF signals are detected and forwarded separately. Therefore $P_{out}$ of the structure with selection at each hop is less than the other structure. It can be seen that at different target $P_{out}$, in both proposed structures, $\gamma_{avg}$ difference between the cases of $M = 1$ and $M = 3$ is more than the cases of $M = 3$ and $M = 5$. It is expected that when M $\to \infty$, system performance reaches a steady state and increasing number of hops does not degrade it any more.

In Fig. 4, Outage Probability of the proposed structures is plotted in terms of average SNR for various variances of Negative Exponential atmospheric turbulence when $M = 2$ and $\gamma_{th} = 10dB$. As can be seen, the structure with selection at each relay has better performance at various variances of Negative Exponential atmospheric turbulence. It can be seen that at high intensity of Negative Exponential atmospheric turbulence ($\lambda = 5$), and different target $P_{out}$, there is slight $\gamma_{avg}$ difference between the proposed structures. In each of the proposed structures, when $P_{out} \leq 10^{-2}$, $\gamma_{avg}$ difference between system performance at various variances of Negative Exponential atmospheric turbulence is fixed. This is a kind of stability, in the sense that addition of a constant fraction of consuming power, maintains performance of the system and there is no need to change this fraction at different situations. This feature eliminates the need for an adaptive processor that changes additional power fraction according to atmospheric turbulences. Therefore, the proposed structures have less complexity, power consumption and processing delay. Supplying the required power is the major issue in the links with saturate atmospheric turbulence regime, hence the proposed structures, due to their properties, are particularly suitable for such links.

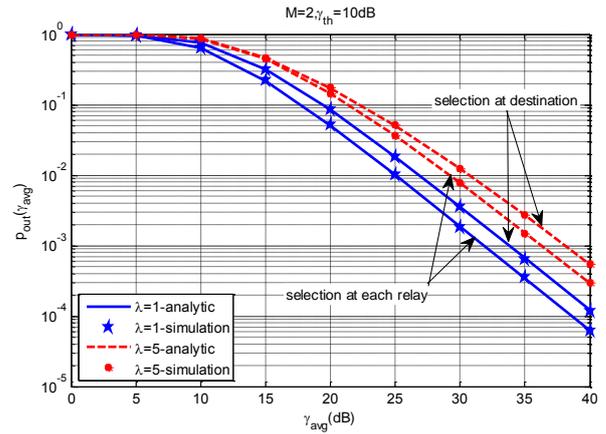

**Fig. 4**: Outage Probability of the proposed structures in terms of average SNR for various variances of Negative Exponential atmospheric turbulence when M = 2 and $\boldsymbol{\gamma_{th} = 10dB}$.

In Fig. 5, Bit Error Rate of the proposed structures is plotted in terms of average SNR for various number of hops for Negative Exponential atmospheric turbulence with unit variance ($\lambda = 1$). As can be seen, at different target $P_{out}$, $\gamma_{avg}$ difference between cases of $M = 2$ and $M = 1$ in the structure with selection at each hop is about 2dB and in the structure with selection at the destination is about 4.5dB, meaning that the structure with selection at the destination consumes about twice (~3dB) more power to maintain the performance. It is also observed that performance difference between the proposed structures increases by adding number of hops. For example, at $P_{out} = 10^{-3}$, $\gamma_{avg}$ difference between the proposed structures in cases of $M = 1$, $M = 2$ and $M =$ 3, are respectively about 0dB, 2dB, and 3dB. Because the structure with selection at the destination makes only one decision, but the other structure makes decision every hop, also detection of each relay is independent from other relays. Therefore, increasing number of hops, increases number of total decisions, thereby increases number of total errors.

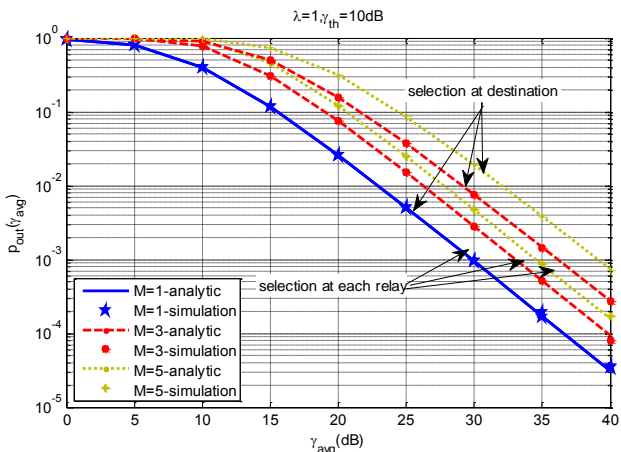

**Fig. 3**: Outage Probability of the proposed structures in terms of average SNR for various number of hops for Negative Exponential atmospheric turbulence with unit variance ($\lambda = 1$) and $\gamma_{th} = 10dB$.

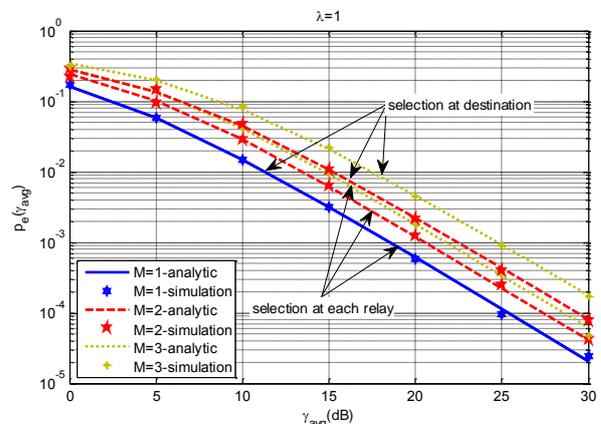

**Fig. 5**: Bit Error Rate of the proposed structures in terms of average SNR for various number of hops, for Negative Exponential atmospheric turbulence with unit variance ($\lambda = 1$).

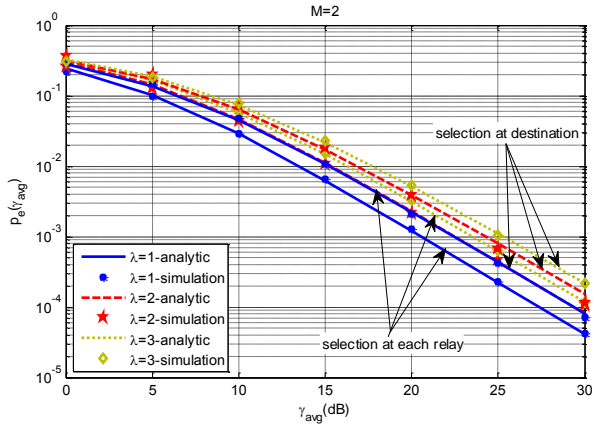

**Fig. 6**: Bit Error Rate of the proposed structures in terms of average SNR for various variances of Negative Exponential atmospheric turbulence when M=2.

In Fig. 6, Bit Error Rate of the proposed structures is plotted in terms of average SNR for various variances of Negative Exponential atmospheric turbulence when M=2. As can be seen, at low $\gamma_{avg}$, the deference between performances of the proposed structures at various atmospheric turbulence variances is not so much, but at high $\gamma_{avg}$ this difference increases. Since at low $\gamma_{avg}$, the noise effect is dominant, i.e., performance degradation is mostly due to the noise effect, but by increase of $\gamma_{avg}$, the effect of atmospheric turbulence overcomes the noise effect. At different target $P_e$, $\gamma_{avg}$ difference between the proposed structures at various atmospheric turbulence variances is the same. When number of hops is constant, in fact number of made decisions is constant, thereby performance difference between proposed structures, which depends on the number of made decisions, is constant and does not change by change in atmospheric turbulence variance.

## 5 Conclusion

In this paper, two novel models are presented for multihop hybrid FSO / RF communication system. In both structures, a series multi-hop relay assisted hybrid FSO / RF link, connects source and destination Base Stations. It is the first time that in a multihop relay assisted hybrid FSO / RF structure, Detect and Forward relaying scheme is used. One of the proposed structures makes decisions at each hop, but the other makes decision only at the last hop. Considering FSO link in Negative Exponential atmospheric turbulence and RF link in Rayleigh fading, for the first time, closed-form expressions are derived for $BER$ and $P_{out}$ of the proposed structures and the derived expressions are verified through MATLAB simulations. It has worth to mention that this is the first time that BER of a multihop hybrid FSO / RF structure is being investigated.

Performance of the proposed structures are compared and investigated at various numbers of hops and different variances of Negative Exponential atmospheric turbulence. It is observed that increasing number of hops degrades performance of both structures. But overall, performance of the first structure is better than the second one, because in the first structure, each relay selects signal with higher SNR for detection but in the second structure, each of the FSO and RF signals are detected and forwarded separately.

Results indicate that at different target $P_{out}$, structure with selection at the destination consumes about two times ($\sim 3dB$) more power than the first structure to maintain performance. At low $\gamma_{avg}$, the difference between performances of the proposed structures at various atmospheric turbulence variances is negligible, therefore the proposed structures are almost independent of atmospheric turbulences and do not need adaptive processor or consuming much more power to maintain performance, thus they have low complexity, power consumption and processing delay. The proposed structures, due to their properties, are particularly suitable for saturate atmospheric turbulence and long-range links, where supplying the required power is the major problem.

## 6 Reference

1   Jamali,V., Michalopoulos, D. S., Uysal, M., Schober, R.: 'Resource Allocation for Mixed RF and Hybrid RF/ FSO Relay', (2015)

2   Lopez-Martinez, F. J., Gomez, G., Garrido-Balsells, J. M.: 'Physical-Layer Security in Free Space Optical Communications', IEEE Photonics Journal 7 (2015)

3   Kaushal, H., Jain, V.K., Ka S.: 'Free Space Optical Communication', Springer India 2017, pp.60

4   Amirabadi, M. A., Vakili, V. T., Amirabadi M., Zamanabadi, K. F.: 'performance of FSO-MIMO communication system at negative exponential atmospheric turbulence channel', 2017 IEEE 4th International Conference on Knowledge-Based Engineering and Innovation, 2017 (In Farsi)

5   Mai, V. V., Pham, A. T.: 'Designs for Multi-Rate Hybrid Adaptive FSO / RF Systems over Fading Channel', Globecom Workshops (2014)

6   Zhang, J., Dai, L., Zhang, Y., Wang Z.: 'Unified Performance Analysis of Mixed Radio Frequency/Free-Space Optical Dual-Hop Transmission Systems', Journal of Lightwave Technology 33 (2015)

7   Jing, Z., Shang-hong, Z., Wei-hu, Z., Ke-fan, C.: 'Performance analysis for mixed FSO/RF Nakagami-m and Exponentiated Weibull dual-hop airborne systems', Optics Communications 392 (2017) 294-299

8   Kumar, K., Borah, D. K.: 'Hybrid FSO/RF Symbol Mappings: Merging High Speed FSO with Low Speed RF through BICM-ID', IEEE Global Communications Conference (2012)

9   Abdulhussein, A., Oka, A., Nguyen, T. T., Lampe, L.: 'Rateless Coding for Hybrid Free-Space Optical and Radio-Frequency Communication', IEEE Transactions on Wireless Communications 9 (2010)

10  Rakia, T., Yang, H.C., Gebali, F. , Alouini, M. S.: 'Power Adaptation Based on Truncated Channel Inversion for Hybrid FSO / RF ,ansmission With Adaptive Combining', IEEE Photonics Journal 7 (2015)

11  Rakia, T., Yang, H. C., Alouini, M. S. , Gebali, F.: 'Outage Analysis of Practical FSO / RF Hybrid System With Adaptive Combining', IEEE Communications Letters 19 (2015)

12  Mai, V. V., Pham, A. T.: 'Adaptive Multi-Rate Designs and Analysis for Hybrid FSO / RF Systems over Fading Channel', IEICE Transactions on Communication e98-B (2015) 1660-1671

13  Zedini, E., Soury, H., Alouini, M. S.: 'Dual-Hop FSO Transmission Systems over Gamma-Gamma Turbulence with Pointing Errors', IEEE Transactions on Wireless Communications, 16 (2017) 784 -796

14  Kumar, N. , Bhatia, V.: 'Performance Analysis of Amplify-and-Forward Cooperative Networks with Best-Relay Selection Over Weibull Channels Fading', Wireless Personal Communications 85 (2015) 641-653

15  Zedini, E., Soury, H., Alouini M. S.: 'On the Analysis of Dual-Hop Mixed Performance FSO/ RF System', IEEE Transactions on Wireless Communications 15 (2016)

16  Djordjevic, G., Petkovic, M., Cvetkovic, A., Karagiannidis, G.: 'Mixed RF/ FSO Relaying with Outdated Channel State Information', IEEE Journal on Selected Areas in Communications PP (2015) 17  Varshney, N., Puri, P.: 'Performance Analysis of Decode-and-Forward MIMO-Based Mixed RF/ FSO Source Mobility and Cooperative Systems with Imperfect CSI', Journal of Lightwave Technology 35 (2017)

18  Saidi, H., Tourki, K., Hamdi, N.: 'Performance analysis of PSK modulation in DF dual-hop hybrid RF / FSO system over gamma gamma channel', International Symposium on Signal, Image, Video and Communications (Isivc) (2016)

19  Chen, L., Wang, W., Zhang, C.: 'Multiuser Diversity Over Parallel and Hybrid FSO / RF Links and Its Performance Analysis', IEEE Photonics Journal 8 (2016)

20  Amirabadi, M. A., Vakili, V. T.: 'A novel hybrid FSO/RF communication system with receive diversity'. arXiv preprint arXiv:1802.07348.

21  Anees, S., Bhatnagar, M.: 'Performance of an Amplify-and-Forward Dual-Hop Asymmetric RF-FSO Communication System', IEEE / OSA Journal of Optical Communications and Networking 7 (2015)

22  Amirabadi, M. A., Vakili, V. T.: 'Performance analysis of a novel hybrid FSO / RF communication system'. arXiv preprint arXiv:1802.07160.

23  Amirabadi, M. A., Vakili, V. T.: 'Performance analysis of hybrid FSO/RF communication systems with Alamouti Coding or Antenna Selection'. arXiv preprint arXiv:1802.07286.

24  Amirabadi, M. A., Vakili, V. T.: 'On the Performance of a CSI Assisted Dual-Hop Asymmetric FSO/RF Communication System over Gamma-Gamma atmospheric turbulence considering the effect of pointing error'. International Congress on Science and Engineering

25  Makki, B., Svensson, T., Brandt-Pearce, M., Alouini, M. S.: 'Performance analysis of RF FSO multi-hop networks', IEEE Wireless Communications and Networking Conference (2017)

26  Kazemi, H., Uysal, M., Touati, F., Haas, H.: 'Outage Performance of Multi-Hop Hybrid FSO / RF Communication Systems', 4th International Workshop on Optical Wireless Communications (2015)

27  Papoulis, A., Pillai, S. U.: 'Probability, Random Variables, and Stochastic Presses', McGraw-Hill, 4th edition 2002, pp.132

28  Kong, L., Xu, W., Hanzo, L., Zhang, H., Zhao, C.: 'Performance of a Free-Space-Optical Relay-Assisted Hybrid RF / FSO System in Generalized M-Distributed Channels', IEEE Photonics Journal 7 (2015)

29  Http: //functions.wolfram.com/